\documentclass[aps,prx,reprint,letterpaper,showpacs,twocolumns]{revtex4-1}
\usepackage{graphicx}
\usepackage{bm}
\usepackage{amsfonts}
\usepackage[colorlinks,urlcolor=blue,anchorcolor=blue,linkcolor=blue,citecolor=blue,breaklinks=true]{hyperref}

\begin{document}

\title{Non-Abelian Fractional Chern Insulator in Disk Geometry}
\author{Ai-Lei He$^{1}$, Wei-Wei Luo$^{2}$, Hong Yao$^{1,3}$, Yi-Fei Wang$^{4}$}
\affiliation{$^1$Institute for Advanced Study, Tsinghua University, Beijing 100084, China
\\$^2$National Laboratory of Solid State Microstructures and Department of Physics, Nanjing University, Nanjing 210093, China
\\$^3$Department of Physics, Stanford University, Stanford, CA 94305, USA
\\$^4$Center for Statistical and Theoretical Condensed Matter Physics, and Department of Physics, Zhejiang Normal University, Jinhua 321004, China}

\date{\today}

\begin{abstract}
  Non-Abelian (NA) fractional topological states with quasi-particles obeying NA braiding statistics have attracted intensive attentions for both its fundamental nature and the prospect for topological quantum computation. To date, there are many models proposed to realize the NA fractional topological states, such as the well-known Moore-Read quantum Hall states and the Non-Abelian fractional Chern insulators (NA-FCIs). Here, we investigate the NA-FCI in disk geometry with three-body hard-core bosons loaded into a topological flat band. This stable $\nu=1$ bosonic NA-FCI is characterized by the edge excitations and the ground-state angular momentum. Based on the generalized Pauli principle and the Jack polynomials, we successfully construct a trial wave function for the NA-FCI. Moreover, a $\nu=1/2$ Abelian FCI state emerges with the increase of the on-site interaction and it can be identified with the help of the trial wave function as well. Our findings not only lead to an optimal wave function for the NA-FCI, but also directly provide an effective approach for future researches on paired topological states.
\end{abstract}

\maketitle

{\it Introduction.---}
One of the most fascinating features in the fractional quantum Hall (FQH) states~\cite{Laughlin,NA1,NA2,NA3,NA4} is that quasi-particles obey the Abelian and non-Abelian braiding statistics. Especially, the non-Abelian fractional quantum Hall (NA-FQH) states~\cite{NA1,NA2,NA3,NA4} have attracted intensive attentions and great interests for its fundamental nature and the prospect of topological quantum computation~\cite{TQC1,TQC2}. One well-known NA-FQH system is the Moore-Read state~\cite{NA1} with the filling factor $\nu=5/2$ which is considered as a p-wave paired state~\cite{Read_Green}. Evidences of Moore-Read state have been experimentally observed~\cite{RE_MR} decades ago, however, the precise nature of this paired state is still in debate. There are many approaches to identifying the Moore-Read state, including the trial wave function (WF)~\cite{NA1}, the edge excitation~\cite{EE1,EE2,EE3,EE4}, the braiding statistics~\cite{NA1,braiding}, the entanglement entropy~\cite{Entangle1,Entangle2}, etc. Among them, the trial WFs provide the key to understanding the nature of NA-FQH states. Fortunately, trial WF of Moore-Read state with a simple analytic expression has been proposed from a conformal field theory perspective~\cite{NA1}. Similar to the Laughlin WFs, the Moore-Read state can be decomposed into anti-symmetric Slater determinants (for fermions) or symmetric monomials (for bosons) with the help of the Jack symmetric polynomials (Jacks)~\cite{Jacks0,Jacks1,Jacks2,Jacks3}, which naturally reflects the generalized Pauli principle (GPP)~\cite{GPP0,GPP1,GPP2,GPP3} in the Fock space. The GPP and the Jacks have been extended~\cite{ALHe,ALHe1} to the lattice analogs of FQH states in the absence of external magnetic field, which are named fractional Chern insulaors (FCIs)~\cite{Sheng1,YFWang1,Regnault1,YFWang2,YFWang3, Tang, Ksun,Neupert, Muller, Qi0,YLWu,YLWu0,Bernevig,YLWu3,ZLiu_NA,ZLiu_NA1,Ronny0,Ronny1,Ronny2,Ronny3,Ronny4,WWLuo,liuz,FCI_reviews,FCI_reviews1}.

Recently, Abelian FCIs have been systematically investigated with interacting particles loaded into the topological flat bands (TFBs)~\cite{Sheng1,YFWang1,Regnault1,YFWang2,YFWang3, Tang, Ksun,Neupert, Muller, FCI_reviews,FCI_reviews1}. Interestingly, there are some proposals to realize the bosonic FQH states in lattice models based on the chiral spin states~\cite{Kalmeyer, Wen3, Ronny5, Yao3, Ronny6, Ronny7}. Meanwhile, various ways to construct the trial WFs for the Abelian FCIs  have been proposed based on the one-to-one mapping relationship~\cite{Qi0} between the FQH states and the FCIs with analytical, semi-analytical~\cite{Qi0,YLWu,YLWu0} or purely numerical approaches~\cite{ALHe,ALHe1}. In terms of the GPP and the Jacks, optimal trial WFs for the Abelian FCIs with interacting particles filling in the TFBs have been constructed successfully~\cite{ALHe,ALHe1}. This direct and effective prescription  has been extended to construct the geometry-dependent $\nu=1/2$ FCI of hard-core bosons~\cite{ALHe1}. Furthermore, it has been used to explore and identify some exotic FCI states~\cite{ALHe11}. Beyond the Abelian FCIs, non-Abelian FCIs (NA-FCIs), especially the $\nu=1/2$ fermionic and $\nu=1$ bosonic NA-FCIs~\cite{YFWang2,Bernevig,YLWu3,ZLiu_NA,liuz,ZLiu_NA1,ZLiu_NA2,WZhu0,WZhu1}, have been studied with interacting particles filling in TFBs. Several numerical studies have provided convincing evidence of these Pfaffian-like NA-FCIs in TFBs with three-body ~\cite{YFWang2,Bernevig,YLWu3,ZLiu_NA,liuz,WZhu0,WZhu1} or long-range two-body interaction~\cite{ZLiu_NA1,ZLiu_NA2}. The NA-FCIs have been investigated on torus geometry which are characterized by quasi-degeneracy of ground states (GSs), many-body Chern number associated with the GSs, robust bulk excitation spectrum gap and even entanglement entropy~\cite{YFWang2,Bernevig,YLWu3,ZLiu_NA,ZLiu_NA1,ZLiu_NA2,WZhu0,WZhu1}. The optimal trial WFs~\cite{ALHe,ALHe1,ALHe11,Qi0,YLWu,YLWu0} and edge excitations~\cite{WWLuo,liuz} are usually considered as a very powerful method to explore the FCIs. However, neither the optimal trial WFs for the NA-FCIs  nor the edge excitations of the NA-FCIs are reported in disk geometry to date.

In this paper, we investigate the NA-FCI in disk geometry with three-body hard-core bosons loaded into a TFB Kagom{\'e} lattice by using the exact diagonalization (ED) method. We find convincing numerical evidence of the stable $\nu=1$ bosonic NA-FCI in disk geometry, characterized by the very clear edge excitation spectra, the degeneracy sequence and the angular momentum of the GS. The quasi-degeneracy sequences of edge excitations for the NA-FCI are directly observed with the ED studies and well predicted by the GPP. Interestingly, different quasi-degeneracy sequences are observed with even and odd boson numbers. The angular momentum of the GS configuration is counted with the help of the GPP as well. Based on the GPP and the Jacks, we construct the trial WFs of the $\nu=1$ bosonic NA-FCI in disk geometry. The high value of the WF overlaps identifies this stable $\nu=1$ bosonic NA-FCI and demonstrates the feasibility of our method. Moreover, we obtain a $\nu=1/2$ Abelian FCI state with the proper on-site repulsive interaction. We systematically explored the Pfaffian-like NA-FCI in disk geometry, and proposed several effective approaches to studying more topological paired states directly.

\begin{figure}[!tp]
\includegraphics[scale=0.9]{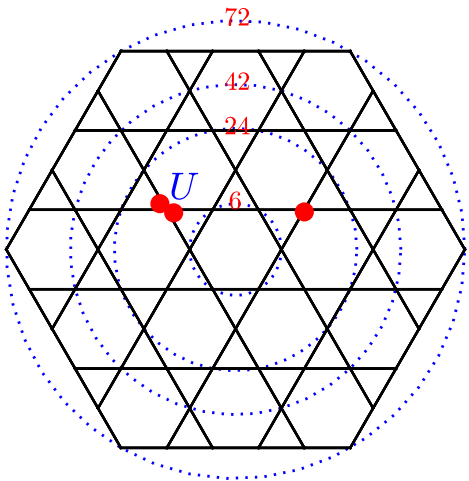}
\caption{(color online). A 72-site Kagom{\'e} disk loaded with three-body hard-core bosons. There are no more than 2 bosons occupying in one site, and the two-body on-site interaction $U$ is added in every sites.}
\label{Model}
\end{figure}

{\it Models.---} We consider the TFB Kagom{\'e}-lattice model with three-body hard-core bosons in disk geometry. The Hamiltonian can be written as~\cite{YFWang2},
\begin{eqnarray}
H= &-&t\sum_{\langle\mathbf{r}\mathbf{r}^{\prime}\rangle}
\left[b^{\dagger}_{\mathbf{r}^{\prime}}b_{\mathbf{r}}\exp\left(i\phi_{\mathbf{r}^{ \prime}\mathbf{r}}\right)+\textrm{H.c.}\right]\nonumber\\
&-&t^{\prime}\sum_{\langle\langle\mathbf{r}\mathbf{r}^{ \prime}\rangle\rangle}
\left[b^{\dagger}_{\mathbf{r}^{\prime}}b_{\mathbf{r}}+\textrm{H.c.}\right] + \frac{U}{2}\sum_{\mathbf{r}}
n_{\mathbf{r}}\left(n_{\mathbf{r}}-1\right)  \label{e.1}
\end{eqnarray}
where $b^{\dagger}_{\mathbf{r}}$ ($b_{\mathbf{r}}$) creates (annihilates) a three-body hard-core
boson at site $\mathbf{r}$ satisfying $\left(b^{\dagger}_{\mathbf{r}}\right)^3=0$ that no more than 2 bosons occupying in any
site are allowed (shown in Fig.~\ref{Model}), and $\left(b_{\mathbf{r}}\right)^3=0$~\cite{YFWang2}. $\langle...\rangle$ and $\langle\langle...\rangle\rangle$ denote the nearest-neighbor (NN), the next-nearest-neighbor (NNN) pairs of sites. Here, we choose the TFB parameters for the Kagom{\'e}-lattice model, i.e. $t=1.0, t^{\prime}=-0.19$ and $\phi=0.22\pi$. $U$ is the two-body on-site interactions (shown in Fig.~\ref{Model}). Clearly, $U/t\to\infty$ corresponds to the two-body hard-core bosons, and a possible 1/2 FCI state may emerge with adding the strength of on-site interaction $U$.

Following the previous works~\cite{ALHe,ALHe1,WWLuo}, an additional trap potential is required in the finite-size systems to obtain the clear edge excitation spectra of FCIs. Here, we choose a harmonic trap with the form $V=V_{\rm{trap}}\sum_{\mathbf{r}}|\mathbf{r}|^2 n_{\mathbf{r}}$ where $V_{\rm trap}$ is the trap potential strength and $|{\mathbf{r}}|$ is the site position distance from the center of the lattice disk~\cite{ALHe,ALHe1,WWLuo}. Root configurations of the stable $\nu=1$ NA-FCI are $|20202...202\rangle$ (for even boson numbers) and $|20202...201\rangle$ (for odd boson numbers). Different from the GS configuration of bosonic $1/2$ FCI $|10101...101\rangle$, there are two root configurations with parity of particle number. These root configurations both obey the GPP. And there is a clear branch of edge excitations. However, the quasi-degeneracy sequences are different.

\begin{figure}[!htb]
\includegraphics[scale=0.8]{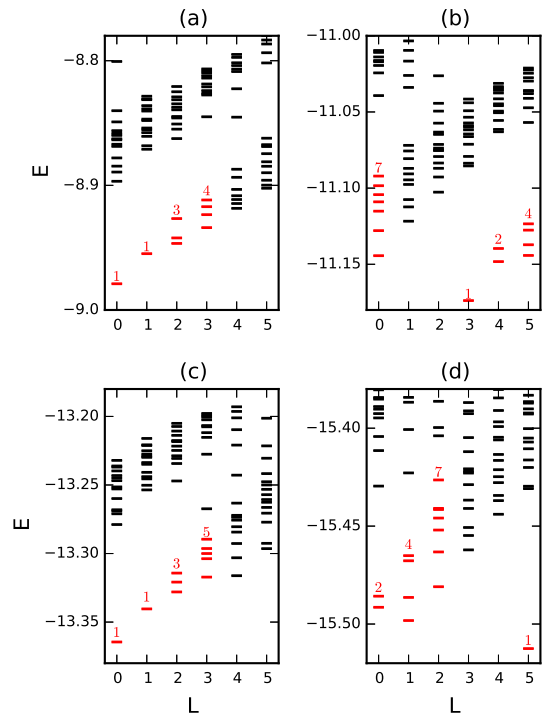}
\caption{(color online). Edge excitations of the $\nu=1$ NA-FCI states in disk geometry with 42 sites and the trap potential $V_{\rm{trap}}=0.0035$. Various numbers of bosons fill into the TFB Kagom{\'e} lattice, and from (a) to (d), the numbers of filling bosons are $N_b=4$, $N_b=5$, $N_b=6$, and $N_b=7$. Numbers labeled upon low-energy levels in each sector show the quasi-degeneracy of low edge excitations.}
\label{EE}
\end{figure}

{\it Edge excitations.---}
The bulk-edge correspondence is a key feature of the FQH states~\cite{EE1,EE2,EE3,EE4} and FCIs~\cite{ALHe,ALHe1,WWLuo,liuz}. The quasi-degeneracies of FQH states are coincident well with the prediction from the chiral Luttinger liquid theory~\cite{EE1,EE2}. Edge excitations for the $\nu=1/m$ FCIs have been directly observed in disk and singular lattices~\cite{ALHe,ALHe1,WWLuo}. Although the number of edge excitation branches is different in disk and singular geometries, the quasi-degeneracy sequences in every branch are the same, i.e. 1,1,2,3,5,7... and configurations of edge excitations in every branch fulfill the GPP that no more than one particle occupy the $m$ consecutive orbitals. Based on the GPP and the Jacks, the edge excitations spectra of FCIs have been estimated in disk geometry~\cite{ALHe}. Unlike edge excitations for the $\nu=1/m$ FQH states, quasi-degeneracy sequences of Moore-Read states are 1,1,3,5,10,... (with even particle numbers) and 1,2,4,7,13,... (with odd particle numbers)~\cite{EE1,EE2}. Edge excitations for Pfaffian state have been obtained by using ED studies~\cite{EE4} and constructed with the aid of the Jacks~\cite{recoEE}.  Moreover, entanglement spectra provide another way to explore the properties of edge excitations for the NA-FQH states and NA-FCIs in torus geometry indirectly~\cite{Entangle2,liuz,WZhu0,WZhu1}. Nevertheless, edge excitations for the NA-FCI in disk geometry have not been studied directly to date.

We investigate edge excitations for these FCIs with various numbers of three-body hard-core bosons filing into the 42-site TFB Kagom{\'e}-lattice disk with the trap potential $V_{\rm{trap}}=0.0035$. One branch of clear edge excitation spectra is observed in our ED results shown in Fig.~\ref{EE} with various numbers of bosons. There are two different quasi-degeneracy sequences,  i.e. 1,1,3,5,... [with the even boson numbers, shown in Fig.~\ref{EE} (a) and (c)] and 1,2,4,7,... [with the odd boson numbers, shown in Fig.~\ref{EE} (b) and (d)] which correspond to the quasi-degenerate sequences of Moore-Read states. In addition, the angular momenta of GS can reflect the information of root configurations which are analyzed from the GPP. We have check the angular momenta of GS with the root configurations $|2020...202\rangle$ (for even boson number cases) and $|2020...201\rangle$ (for odd boson number cases). They are in good accordance with our ED results (more details shown in Supplement.~\ref{QDS}). Based on the GPP, the root configurations of edge excitations in every sector have been counted, and the results correspond with the quasi-degeneracy of our ED results (more details shown in Supplement.~\ref{QDS}). Heuristically, we conjecture that these FCIs are the bosonic $\nu=1$ NA-FCI. Following the previous works~\cite{ALHe,ALHe1,ALHe11}, optimal trial WFs can be used to identify the FCIs. To verify our conjecture, we construct the trial WF for the NA-FCI.

{\it Optimal trial wave functions.---}
Base on the GPP and the Jacks, optimal trial WFs for FCIs have been constructed in disk and singular geometries~\cite{ALHe,ALHe1,ALHe11} which can be used to identify different FCI states. According to the quasi-degeneracy sequences of the edge excitations and the angular momenta of the GS, we conjecture that these exotic FCI states are the bosonic $\nu=1$ NA-FCI which can be identified by the optimal trial WFs. Here, we first consider the WFs of bosonic $\nu=1$ Moore-Read state in an infinite-size disk with complex positions $\{z_i\}$ as~\cite{NA1},
\begin{equation}\label{MRW}
\Psi_{\rm{MR}}(\{z_i\})={\rm{Pf}}(\frac{1}{z_i-z_j})\prod_{i<j}(z_i-z_j){\rm{e}}^{{-\sum_{i} \frac{|z_i|^2}{4}}}.
\end{equation}
Here, the Pfaffian (Pf) is defined by
\begin{equation}\label{PF}
{\rm{Pf}}M_{ij} = \frac{1}{2^N(N/2)!}\sum_{\sigma} {\rm{sgn}}\sigma \prod_{k=1}^{N/2} M_{\sigma(2k-1)\sigma(2k)}.
\end{equation}
$M$ is an $N\times N$ antisymmetric matrix with the element $M_{ij}$. ${\rm{Pf}}(\frac{1}{z_i-z_j})$ in Eq.~\ref{MRW} is equal to ${\cal{A}}(\frac{1}{z_1-z_2}\frac{1}{z_3-z_4}...)$, where $\cal{A}$ is the anti-symmetric operation (we show the specific form in Supplement.~\ref{MR_EX} with finite number of bosons). Similar to the Laughlin states, this bosonic Moore-Read state can be decompose into symmetric monomials (the Jacks basis) based on the Jacks~\cite{Jacks0,Jacks1,Jacks2,Jacks3}, i.e. the normalized $\Psi^{N}_{\rm{MR}}(\{z_i\})=\sum_kc^{N}_k\Psi^{J}_k(\{z_i\})$. $\Psi^{J}_k(\{z_i\})$ is the Jacks basis and $|c^{N}_k|^2$ is the probability of the $k$th Jacks basis (more details shown in Supplement.~\ref{MR_EX}).

There is a mapping relationship between the FQH states and the FCIs, and bsaed on this mapping relationship~\cite{Qi0}, trial WFs for $\nu=1/m$ FCI have been constructed in the cylinder, torus , disk and even singular geometries~\cite{Qi0,YLWu,YLWu0,ALHe,ALHe1}. However, the optimal trial WF for the NA-FCI has not been studied in disk geometry. Here, we make the bold conjecture that the mapping relationship between the NA-FQH states and the NA-FCIs still exist and the optimal trial WF for the $\nu=1$ NA-FCI $\Psi_{{\rm FCI}}^{\nu=1}(\{z_i\})$ can be constructed based on the GPP, the Jacks and the single-particle states$\{\psi_m\}$. We first construct the soft-core bosonic FCI WF $\Psi_{{\rm SCB}}^{\nu=1}(\{z_i\})$ with the GS root configuration $|202020...0202\rangle$ (even boson numbers) and $|202020..0201\rangle$ (odd boson numbers), i.e. $\Psi_{{\rm SCB}}^{\nu=1}(\{z_i\})=\sum_{k}c^{N}_k\Psi_k(\{z_i\})$, where $\Psi_k(\{z_i\})$ is a many-particle WF with free bosons and $|c^{N}_k|^2$ is the proportion of the $k$th Jacks basis. In order to reflect the peculiarity of  three-body hard-core bosons, the trial WFs are constructed with the help of the effective projection operator,
\begin{equation}\label{PJ}
\Psi^{\nu=1}_{\rm FCI}={\cal N}_{\rm FCI} \widehat{\cal P} \Psi_{\rm SCB}.
\end{equation}
Here, $\widehat{\cal P}=\prod_{i,j,k}(1-\delta_{i,j}\delta_{j,k}\delta_{k,i})$ is a projection operator which more than two particles are not allowed to occupy in the same site with i, j, k marking the site position. $\delta_{i,j}$ is the Kronecker delta function. The trial WFs can be constructed after normalizing the projection WFs with a normalization coefficient ${\cal N}_{\rm FCI}$.

\begin{table}
\begin{tabular} {c c c c}
 \hline\hline
$N_b$~& ~${\cal D}_{\rm ED}$  ~& ~${\cal D}_{\rm Jack}$ ~& ~${\cal O}$ \\
\hline
3 &  $C^3_{42}+C^1_{42}C^1_{41}$ (13202)  & 2 & 0.978  \\
4 &  $C^4_{42}+C^1_{42}C^2_{41}+C^2_{42}$ (147231) & 3 & 0.960 \\
5 &  $C^5_{42}+C^1_{42}C^3_{41}+C^2_{42}C^1_{40}$ (1332828) & 9 & 0.958 \\
6 &  $C^6_{42}+C^1_{42}C^4_{41}+C^2_{42}C^2_{40}+C^3_{42}$ (10182186) & 16 & 0.944 \\
 \hline
\end{tabular}
\caption{Overlap between the optimal Jack WFs and the ED WFs of the $\nu=1$ FCI. The values of the overlap for the GS (${\cal O}$) are listed for different boson numbers $N_b$. We compare the dimensions between the ED and the Jacks, and ${\cal D}_{\rm ED}$ and ${\cal D}_{\rm Jack}$ denote the dimensions of the ED and the Jack WFs.}
\label{overlap}
\end{table}

We compare the trial WF of this $\nu=1$ FCI and the ED result through calculation of the WF overlap, i.e. $ {\cal O} = |\langle\Psi_{{\rm{ED}}}(\{z_i\})|\Psi^{\nu=1}_{{\rm FCI}}(\{z_i\})\rangle|$. Values of WF overlap are 0.978, 0.960, 0.958 and 0.944 for the
NA-FCI systems with $N_b=3, 4, 5,$ and 6 bosons shown in Table.~\ref{overlap}. Here, a 42-site Kagom{\'e} disk with the trap potential $V_{\rm{trap}}=0.005$ is chosen. The WF overlap is higher than 0.94, even the dimension of Hilbert space is more than $10^7$. The FCI with three-body hard-core bosons is indeed the stable $\nu=1$ NA-FCI according to the high values of the WF overlap.
With the help of the GPP and the Jacks, our approach can not only construct the optimal trial WF for the NA-FCI successfully, but also reduce the dimension of Fock space greatly. The dimension of Hilbert space with ED studies is ${\cal D}_{\rm ED}=\sum_i C^{n_{2,i}}_{N_s}C^{n_{1,i}}_{N_s-n_{2,i}}$ in a $N_s$-site disk geometry. Here, $n_{1,i}$ and $n_{2,i}$ denote the number of lattice sites filled with one and two bosons, and the number of total bosons is $N_b= n_{1,i}+2n_{2,i}$. We have shown these dimension ${\cal D}_{\rm ED}$ in Table.~\ref{overlap} (with the max dimension more than $10^7$). However, the max dimension of the Jacks basis (with 6 bosons) is only 15 which is less than the dimension of ED results. The trial WFs for more than 6 bosons can be obtained with the help of the GPP and the Jacks. Meanwhile, the real-space density $\rho(\bf{r})$ of the NA-FCI is predicted based on the occupation density in the $m$th orbital $\rho_m$ and the single-particle states $\{\phi_m(\bf{r})\}$ (more details shown in Supplement.~\ref{Denpr}).

\begin{figure}[!htb]
\includegraphics[scale=0.65]{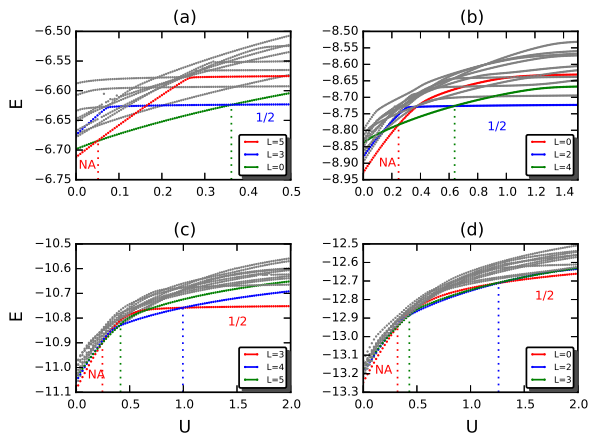}
\caption{(color online). The emergence of possible 1/2 FCI state with the increased $U$ for the 42-site disk with 3 (a), 4 (b), 5 (c) and 6 (d) bosons with the trap potential $V_{\rm{trap}}=0.005$. The angular momentum (L) of every GS energy has been marked. We plot some vertical lines through the crossing point of GS energy. ``NA" and ``1/2" denote the $\nu=1$ NA-FCI and the $\nu=1/2$ FCI. }
\label{PHT}
\end{figure}

{\it 1/2 FCI state.---}
The $\nu=1$ NA-FCI has been observed in the TFB models with three-body hard-core bosons and without the on-site interaction $U$, i.e. $U=0$. When the on-site interaction $U$ is chosen as $U/t\to \infty$, two bosons can not occupy in the same sites which corresponds to the two-body hard-core interaction.  Following the previous works~\cite{YFWang1,WWLuo}, a bosonic 1/2 FCI has been observed only with two-body hard-core bosons. Therefore, the 1/2 Abelian FCI state may be observed by tuning the on-site interaction $U$. With the increase of $U$, the GS energy crossing emerges (shown in Fig.~\ref{PHT}) which implies that the possible 1/2 FCI state can be obtained.

We choose a state far away from the NA-FCI region and check its angular momentum of GS. For example, based on the GPP, the GS angular momentum of a 1/2 FCI with 4 bosons is ``$L$ = (5+1+3+5) $mod$ 6 = 2" (with root configuration $|1010101\rangle$)  and we find that the GS angular momentum of this state in the last region of Fig.~\ref{PHT} (b) is in good accordance with the 1/2 FCI state. Other situations with different numbers of bosons have been checked as well. We conjecture that this state in the last region of Fig.~\ref{PHT} is a 1/2 FCI state.  There is a clear branch of edge excitations with quasi-degeneracy ``1,1,2,3..." (more details shown in Supplement.~\ref{LEE0}) of this state. This state was further completely identified as a 1/2 FCI state with the help of its optimal trial WF. This trial WF of 1/2 FCI state can be constructed based on the GPP and the Jacks~\cite{ALHe,ALHe1} as well. Values of the WF overlap are ${\cal O}(N_b=3,U=0.45)=0.988$, ${\cal O}(N_b=4,U=1.0)=0.991$, ${\cal O}(N_b=5,U=1.5)=0.966$ with $N_b$ bosons and various on-site interactions $U$. This state is indeed the 1/2 bosonic FCI. We have also shown the NA-FCI and 1/2 FCI WFs overlap with tuning the on-site interaction $U$ (more details shown in Supplement.~\ref{LEE0}). From our ED study, we obtain a remarkably robust $\nu=1$ NA-FCI state, and meanwhile, by tuning the on-site interaction $U$, the 1/2 FCI state is observed.  However, some uncertain intermediate states emerge between the NA-FCI state and the 1/2 FCI state shown in Fig.~\ref{PHT}. There is no clear edge excitations of these states (more details shown in Supplement.~\ref{LEE0}). These uncertain intermediate states remain to be further discussed.

{\it Summary and discussion.---}
We have studied the FCI with three-body hard-core bosons filling into the TFB Kagom{\'e} lattice in disk geometry. With the ED results, a clear branch of edge excitations is observed and the quasi-degeneracy sequences with particle parity correspond to the Moore-Read states, i.e. 1,1,3,5,10...for even boson numbers and 1,2,4,7,13...for odd boson numbers. This FCI is identified by the GS angular momenta and the trial WF. The optimal trial WF can be constructed based on the GPP and the Jacks. On one hand, the FCI states are viewed as the stable $\nu=1$ NA-FCI with the high values of WFs overlap, on the other hand, a direct and effective approach are proposed to explore the NA-FCI. Furthermore, a 1/2 Abelian FCI state emerges with the increase of the on-site repulsive interaction. Our findings can open up several future researches on FCIs. How to characterize the uncertain intermediate states between the NA-FCI and the 1/2 FCI in Fig.~\ref{PHT} and identify them. Other exotic paired states of FCIs have been rarely reported, such as the Gaffnian-like FCIs and Parafermion-type FCIs. Geometry-dependent FQH ~\cite{Geo0} and Abelian FCI~\cite{ALHe1} states have been proposed in curved geometries~\cite{ALHe2}, however, geometrical description of NA-FCIs and other paired FCIs are not considered to date. In addition, the anti-Pfaffian states~\cite{ATPF1,ATPF2} have attracted more interests and these anti-Pfaffian-like states are expected to be realized in FCIs.

{\it Acknowledgments.---}
This work is supported by the NSFC of China Grants Nos. 11874325 (YFW) and 11825404 (ALH and HY), the MOSTC under Grant Nos. 2016YFA0301001 and 2018YFA0305604 (HY), the Strategic Priority Research Program of Chinese Academy of Sciences under Grant No. XDB28000000 (HY), Beijing Municipal Science \& Technology Commission under Grant No. Z181100004218001 (H.Y.), and Beijing Natural Science Foundation under Grant No. Z180010 (H.Y.).

\bibliography{NA_FCIs}

\newpage

\section*{SUPPLEMENTARY MATERIALs FOR ``Non-Abelian Fractional Chern Insulator in Disk Geometry''}
\setcounter{figure}{0}
\setcounter{equation}{0}
\renewcommand \thefigure{S\arabic{figure}}
\renewcommand \theequation{S\arabic{equation}}
\setcounter{section}{0}
\renewcommand \thesection{S\arabic{section}}
In the main text, we study fractional Chern insulators (FCIs) with three-body hard-core bosons in disk geometry with topological flat band (TFB) parameters. These FCI states are the stable $\nu=1$ non-Abelian FCI (NA-FCI) which are characterized by the edge excitations and the ground-state (GS) angular momentum.
We construct the optimal trial wave function (WF) for the NA-FCI and high WF overlap values show the stability of the NA-FCI and the feasibility of our trial WFs. Furthermore, a 1/2 Abelian FCI state emerges with the increase of the on-site interaction $U$. In this part, we will show some details about the Moore-Read states, the explanation of the edge excitations, counting the GS angular momenta, the density profile with more bosons and other way to distinguish the $\nu=1$ NA-FCI and the $\nu=1/2$ FCI.

\section{Moore-Read states and its expansion} \label{MR_EX}
The WF for the Moore-Read state has been shown in Eq.~\ref{MRW} with the Pfaffian algebra. The Pfaffian can be denoted as  the anti-symmetric operation, i.e. ${\rm{Pf}}(\frac{1}{z_i-z_j})$ = ${\cal{A}}(\frac{1}{z_1-z_2}\frac{1}{z_3-z_4}...)$. Firstly, we take four bosons as an example, $\Psi_{\rm{MR}}(z_1,z_2,z_3,z_4)={\cal{A}}(\frac{1}{z_1-z_2}\frac{1}{z_3-z_4})[(z_1-z_2)(z_1-z_3)(z_1-z_4)(z_2-z_3)(z_2-z_4)(z_3-z_4)]$ and ${\cal{A}}(\frac{1}{z_1-z_2}\frac{1}{z_3-z_4})=\frac{1}{z_1-z_2}\frac{1}{z_3-z_4}-\frac{1}{z_1-z_3}\frac{1}{z_2-z_4}+\frac{1}{z_1-z_4}\frac{1}{z_2-z_3}$. The Moore-Read state can be expanded as, $\Psi^{(4)}_{\rm{MR}}(z_1,z_2,z_3,z_4)=+(1)(z^2_1z^2_2z^0_3z^0_4+z^2_1z^2_3z^0_2z^0_4+z^2_1z^2_4z^0_2z^0_3+z^0_1z^0_3z^2_2z^2_3+z^0_1z^0_3z^2_2z^2_4+z^0_1z^0_2z^2_3z^2_4)
+(-1)(z^2_1z^1_2z^1_3z^0_4+z^2_1z^1_2z^0_3z^1_4+z^2_1z^0_2z^1_3z^1_4+z^2_2z^1_1z^1_3z^0_4+z^1_1z^2_2z^0_3z^1_4
+z^0_1z^2_2z^1_3z^1_4+z^0_4z^2_3z^1_1z^1_2+z^2_3z^1_1z^1_4z^0_2+z^0_1z^2_3z^1_2z^1_4+z^2_4z^1_1z^2_2z^0_3+z^2_4z^1_1z^1_3z^0_2+z^0_1z^2_4z^1_2z^1_3)+
(+6)z^1_1z^1_2z^1_3z^1_4$. This can be rewritten as $\Psi^{(4)}_{\rm{MR}}=(+1)\Psi_{202}+(-1)\Psi_{121}+(+6)\Psi_{040}$ by using root configuration $|202\rangle$,  here, $\Psi_{202}=z^2_1z^2_2z^0_3z^0_4+z^2_1z^2_3z^0_2z^0_4+z^2_1z^2_4z^0_2z^0_3+z^0_1z^0_3z^2_2z^2_3+z^0_1z^0_3z^2_2z^2_4+z^0_1z^0_2z^2_3z^2_4$, $\Psi_{121}=z^2_1z^1_2z^1_3z^0_4+z^2_1z^1_2z^0_3z^1_4+z^2_1z^0_2z^1_3z^1_4+z^2_2z^1_1z^1_3z^0_4+z^1_1z^2_2z^0_3z^1_4
+z^0_1z^2_2z^1_3z^1_4+z^0_4z^2_3z^1_1z^1_2+z^2_3z^1_1z^1_4z^0_2+z^0_1z^2_3z^1_2z^1_4+z^2_4z^1_1z^2_2z^0_3+z^2_4z^1_1z^1_3z^0_2+z^0_1z^2_4z^1_2z^1_3$ and $\Psi_{040}=z^1_1z^1_2z^1_3z^1_4$. Then we consider the 3-boson Moore-Read WFs. $\Psi^{(3)}_{\rm{MR}}(z_1,z_2,z_3)=(\frac{1}{z_1-z_2}-\frac{1}{z_1-z_3}+\frac{1}{z_2-z_3})[(z_1-z_2)(z_1-z_3)(z_2-z_3)]=(+1)(z^2_1z^0_2z^0_3+z^0_1z^2_2z^0_3+
z^0_1z^0_2z^2_3)+(-1)(z^1_1z^1_2z^0_3+z^1_1z^0_2z^1_3+z^0_1z^1_2z^1_3)$. And it can be write as $\Psi^{(3)}_{\rm{MR}}=(+1)\Psi_{201}+(-1)\Psi_{120}$. However, expanding the Moore-Read state with more than 4 particles is very difficult. Fortunately, the Jacks provide an effective way to decompose Moore-Read state into symmetric monomials (for bosons) or
anti-symmetric Slater determinants (for fermions)~\cite{Jacks0,Jacks1,Jacks2,Jacks3}, i.e. the non-normalized $\Psi_{\rm{MR}}(\{z_i\})=\sum_kc_k\Psi^{J}_k(\{z_i\})$. These basis states $\Psi^{J}_k(\{z_i\})$ and the expansion coefficients $c_k$ can be obtained with the aid of ``squeezing rule" and a recurrence relation~\cite{Jacks0,Jacks1,Jacks2,Jacks3}. The Moore-Read state can be normalized $\Psi^{N}_{\rm{MR}}(\{z_i\})=\sum_kc^{N}_k\Psi^{J}_k(\{z_i\})$ with the normalized coefficients $c^{N}_k$, and $\sum_k|c^{N}_k|^2=1$.

\begin{figure}[!tb]
\includegraphics[scale=0.7]{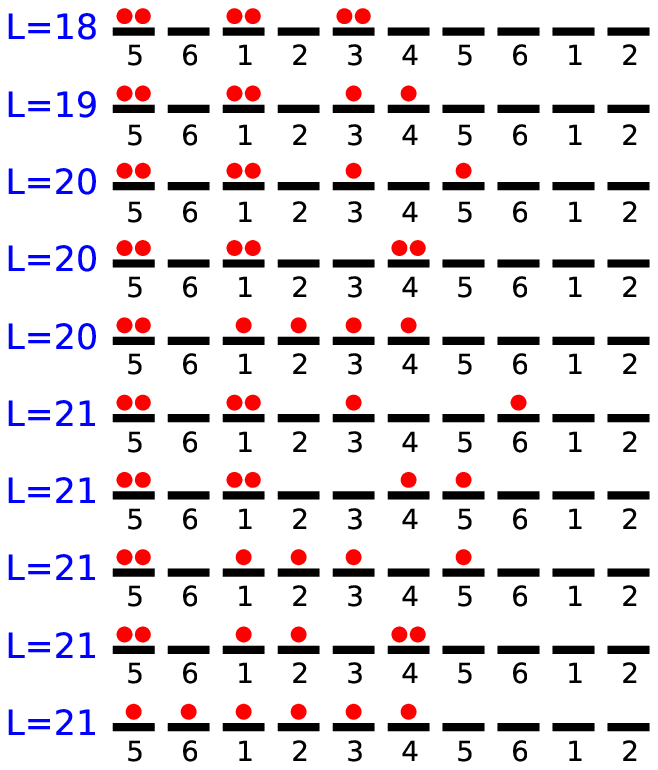}
\caption{(color online). Root configurations of the low-energy excitation modes with 6 bosons filling into TFB orbitals.}
\label{RCE}
\end{figure}

\begin{figure}[!tb]
\includegraphics[scale=0.7]{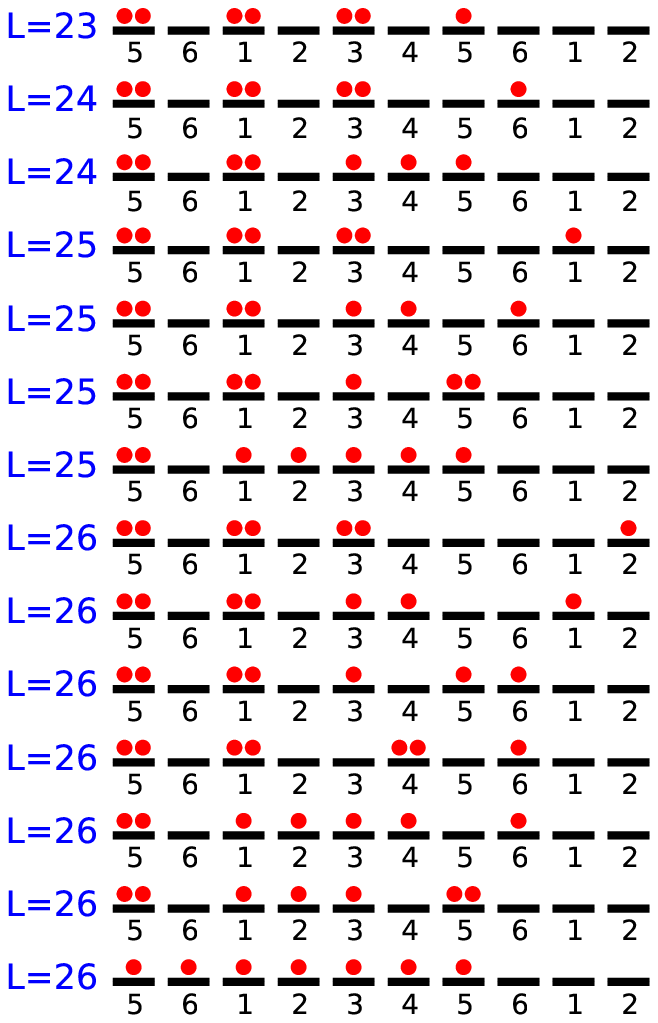}
\caption{(color online). Root configurations of the low-energy excitation modes with 7 bosons filling into TFB orbitals.}
\label{RCO}
\end{figure}

\section{Counting the quasi-degeneracy sequences} \label{QDS}
The edge excitations of $1/2$ and $1/3$ FCIs have been observed in disk geometry with the same quasi-degeneracy sequences, 1,1,2,3,5,7,11...~\cite{WWLuo,ALHe}. And these quasi-degeneracy sequences can be explained in view of the Jacks~\cite{WWLuo}. Here, we show how to count the quasi-degeneracy sequences with the Jacks and we list the root configurations of the low-energy excitation modes based on the generalized Pauli principle (GPP). For a TFB Chern insulator with adding the trap potential in Kagom{\'e} disk, the angular momentum is 5~\cite{WWLuo,ALHe,ALHe1}. We consider 6 and 7 bosons filling into TFB orbitals, and the root configurations of the low-energy excitations are shown in Fig.~\ref{RCE} and Fig.~\ref{RCO}. The root configurations of edge excitations can be denoted with ``0", ``1" and ``2", which ``0" means no bosons, ``1" denotes that only one boson occupies in an orbital and ``2" denotes that two bosons occupies in an orbital. For the 6 bosons filling into TFB orbials, the root configuration of GS is $|20202\rangle$ with total angular momentum 18(mod 6) which is fully consistent with the result shown in Fig.~\ref{EE} (c). Based on the GPP, root configurations of several excited states have been listed in Fig.~\ref{RCE}, and the quasi-degeneracy sequence is 1,1,3,5... which corresponds to the edge excitations of the Moore-Read state with even particle numbers.

Another case which we have considered is the $\nu=1$ NA-FCI filling with odd particle numbers. Here, we take 7 bosons as an example. The root configuration of the GS is $|2020201\rangle$. Considering bosons fill into the TFB orbitals with the first angular momentum 5, the total angular momentum of this NA-FCI is 23 (mod 6), which is as well fully consistent with the result shown in Fig.~\ref{EE} (d). Part of root configurations of excited states can be counted with the aid of the GPP (shown in Fog.~\ref{RCO}) and the quasi-degeneracy sequence is 1,2,4,7, .... These Moore-Read-like states with odd particle numbers can be viewed as a combination of the Pfaffian-like states with even particle numbers and the Laughlin-like states. The quasi-degeneracy sequence is ``\{1,1,3,5,10,...\}+\{0,1,1,2,3,...\}", i.e. ``1,2,4,7,13...".

\begin{figure}[!tb]
\includegraphics[scale=0.42]{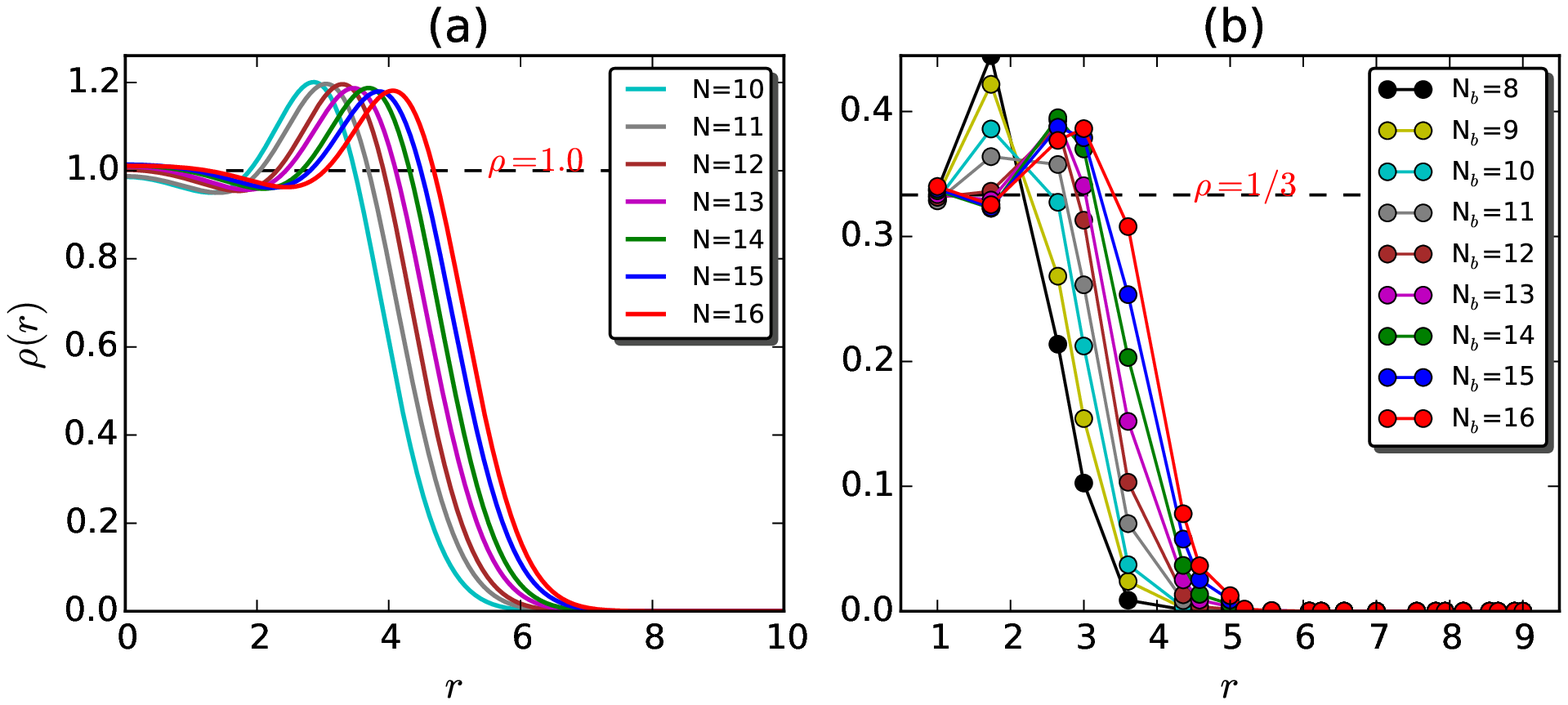}
\caption{(color online). The real-space density profile of the Moore-Read states and the $\nu=1$ NA-FCI states. $r$ is the distance from the center of the circle.}
\label{Dens}
\end{figure}

\section{Density profile of the $\nu=1$ NA-FCI} \label{Denpr}
Moore-Read state can be expanded based on the Jacks, and the real-space density profile $\rho(r)$ can be written with the help of the Jacks, i.e. $\rho(r)=\sum_m\frac{1}{2^mm!}r^{2m}\exp(-\frac{r^2}{2})\rho_m$, and here $\rho_m$ denotes the occupation density in the $m$th orbital. The real-space density of Moore-Read states has shown in Fig.~\ref{Dens} with 10-16 bosons filling in Landau levels.  In the main text, we have show the ED results with 7 bosons, however, the NA-FCI with more than 7 bosons can not be obtained easily by using ED studies. Fortunately, we provide a direct and effective method to construct the optimal trial WFs with the GPP and the Jacks. Meanwhile, inspired by the real-space density profile of Moore-Read state, the density of the $\nu=1$ NA-FCI can be obtained as $\rho(r)=\sum_m |\phi_m(r)|^2\rho_m$ with the single particle states $\{\phi_m\}$.  Here, we can obtain the density profile up to 16 bosons shown in Fig.~\ref{Dens}(b). The Kagom{\'e} lattice has three energy bands, and interaction bosons only occupy in the lowest band (i.e.the TFB), thus the particle density of the $\nu=1$ NA-FCI state is about $1/3$ near the center of the disk.

\begin{figure}[!htb]
\includegraphics[scale=0.5]{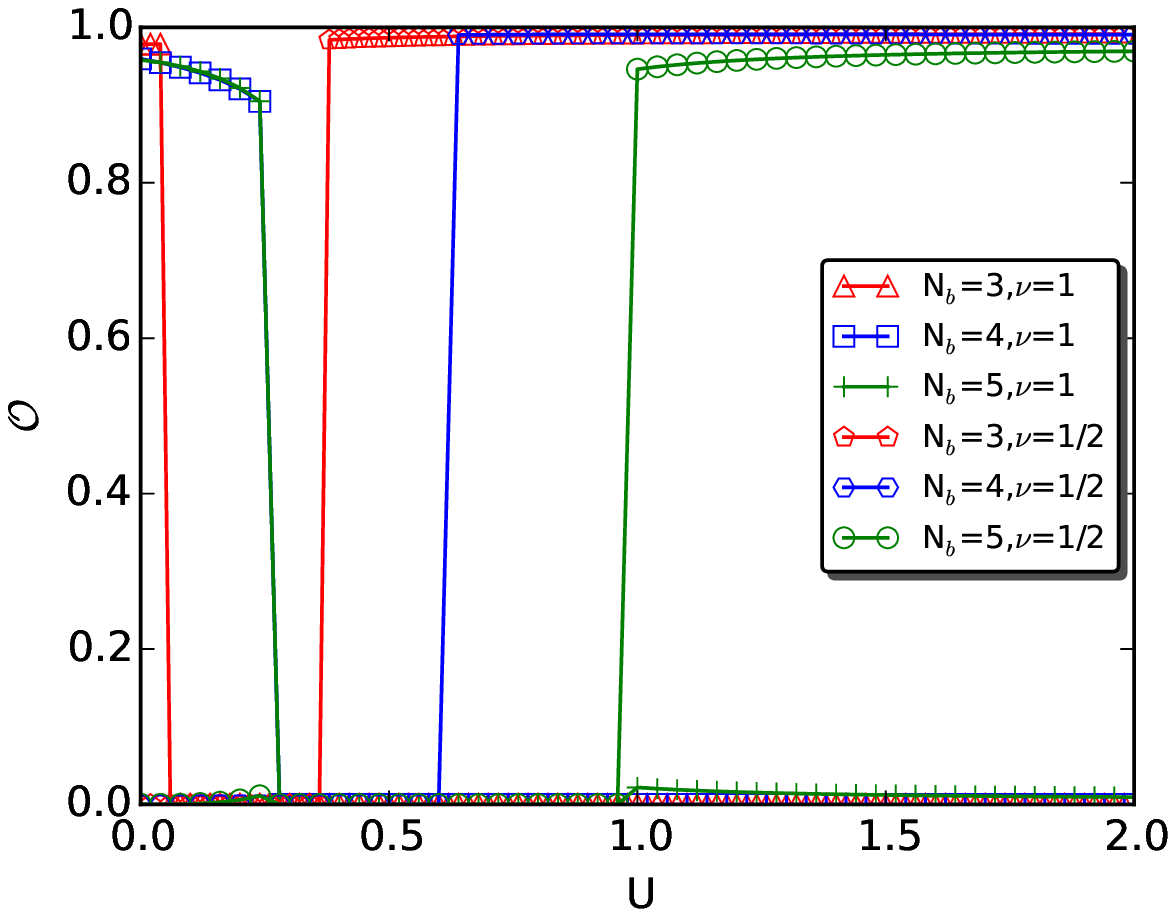}
\caption{(color online). Values of WF overlap ($\cal{O}$) between the trial WFs and ED results with tuning the on-site interaction U. }
\label{wf_dens}
\end{figure}

\section{Wave-function overlap and low-energy excitations of various states} \label{LEE0}
\begin{figure}[!htb]
\includegraphics[scale=0.7]{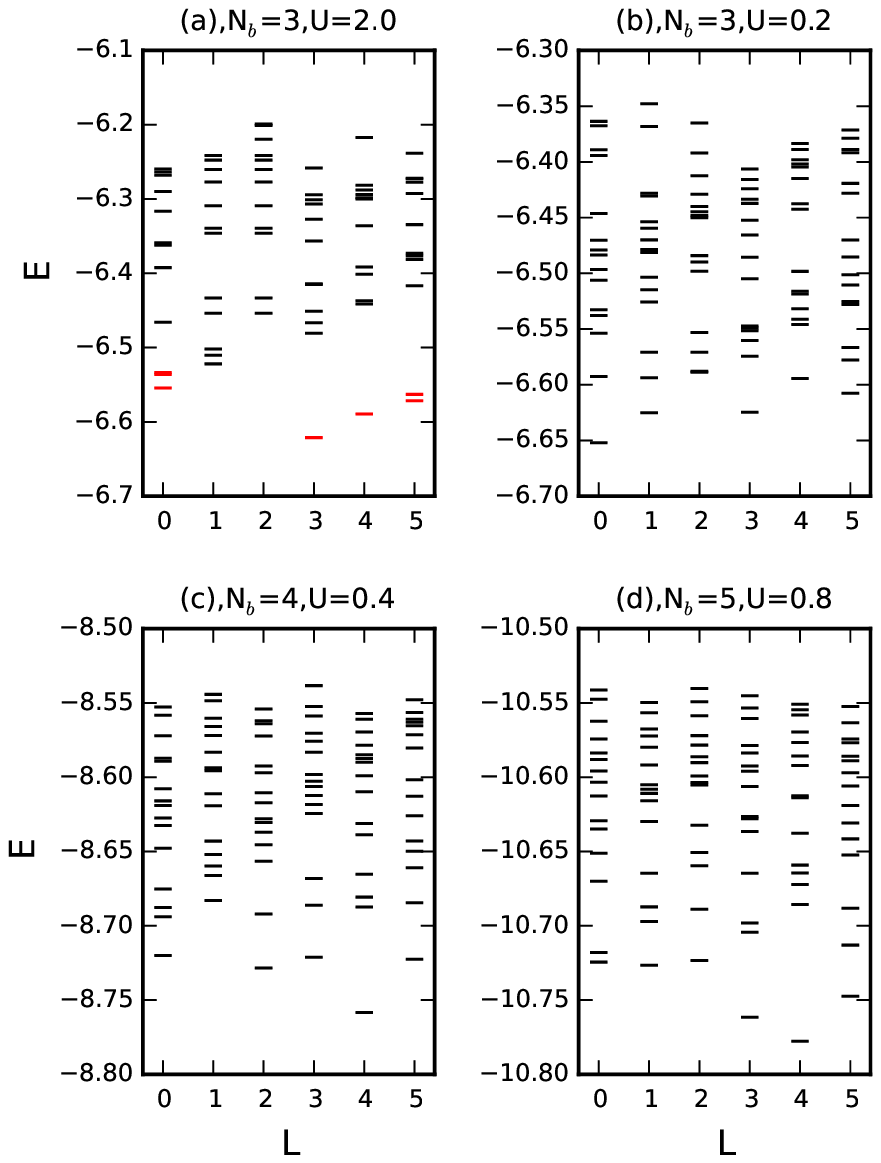}
\caption{(color online). Low-energy excitations of states with different interaction $U$. (a). Edge excitations of the FCI states with quasi-degeneracy ``1,1,2,3...". (b)-(c) Low-energy excitations of uncertain intermediate states.}
\label{LEE}
\end{figure}

FCIs with various filling factors can be revealed in view of the trial wave function (WF) as well. We have shown the values of $\nu=1$ and $\nu=1/2$ WF overlap with tuning the onsite interaction $U$ in Fig.~\ref{wf_dens} (with 3-, 4- and 5- bosons). Here, high values of WF overlap can identify the $\nu=1$ NA-FCI and the Abelian $\nu=1/2$ FCI. As we all know, the 1/2 FCI has a clear branch of energy excitations with a quasi-degeneracy order ``1,1,2,3...". When the on-site interaction is large enough, an ordered edge excitation emerge with three-body bosons loaded into the TFB (shown in Fig.~\ref{LEE} (a)) which is characterized by the 1/2 FCI. With the aid of the trail WFs, we identified this phase as the 1/2 FCI.  However, between the $\nu=1$ NA-FCI region and the 1/2 FCI region, there are some uncertain intermediate states without clear branches of edge excitations (shown in Fig.~\ref{LEE} (b)-(d)). These exotic states remain to be further discussed.

\end{document}